\documentstyle[preprint,aps]{revtex}

\begin{document}
\draft
\title{Why is nacre strong? II: remaining mechanical weakness for cracks
propagating along the sheets}
\author{Shortened title: Why is nacre strong? II: remaining mechanical
weakness}
\author{Ko Okumura\thanks{%
Permanent address: Departement of Physics, Graduate School of Humanities and
Sciences, Ochanomizu University, 2-1-1, Otsuka, Bunkyo-ku, 112-8610, Japan.
e-mail: okumura@phys.ocha.ac.jp}}
\address{Physique de la Mati\`{e}re Condens\'{e}e, Coll\`{e}ge de France,\\
11, place Marcelin-Berthelot, 75231 Paris cedex 05, France}
\date{\today}
\maketitle

\begin{abstract}
In our previous paper (Eur. Phys. J. E 4, 121 (2001)) we proposed a
coarse-grained elastic energy for nacre, or stratified structure of hard and
soft layers. We then analyzed a crack running perpendicular to the layers
and suggested one possible reason for the enhanced toughness of this
substance. In the present paper, we consider a crack running parallel to the
layers. We propose a new term added to the previous elastic energy, which is
associated with the bending of layers. We show that there are two regimes
for the parallel-fracture solution of this elastic energy; near the fracture
tip the deformation field is governed by a parabolic differential equation
while the field away from the tip follows the usual elliptic equation.
Analytical results show that the fracture tip is lenticular, as suggested in
a paper on a smectic liquid crystal (P. G. de Gennes, Europhys. Lett. 13
(8), 709 (1990)). On the contrary, away from the tip, the stress and
deformation distribution recover the usual singular behaviors ($\sqrt{x}$
and $1/\sqrt{x}$, respectively, where $x$ is the distance from the tip).
This indicates there is no enhancement in toughness in the case of parallel
fracture.
\end{abstract}


\preprint{Coll\`{e}ge de France}


\section{Introduction}

A number of {\em living soft matters} derive their strength from some
composite structures. Tooth and timber are among the examples. These
composite structures have motivated studies mainly in industry or
technology-oriented field to produce strong materials, such as raw material
for construction or air plane and automobile tires. \cite%
{Kelly,Clegg,Composites} One of the purposes of this paper is to present
some complementary understandings of such a problem from a viewpoint of a
physicist.

In the first of this series of papers, \cite{KOPGG} which deepened the
scaling picture presented in \cite{PGGKO}, we studied nacre \cite%
{Currey,Jackson,Sarikaya,Rao} which has an alternating laminated structure
of hard and soft layers (Fig. 1). One of the significant features of this
substance lies in the length scale of the layers. The thicknesses of hard
inorganic layer $d_{h}$ and that of the soft organic layer $d_{s}$ are of
the order of nanometer and micron, respectively. This allows us a
coarse-grained treatment in which {\em average deformation field} may be
well defined and can be used for the analysis of the field {\em even near
the tip}. Simplifying the coarse-grained elastic energy further in terms of
Fourier components of the deformation field, which is somewhat different
from conventional approaches, \cite{Composites} we could show analytical
solutions and thereby demonstrate one possible mechanism for the toughness
of nacre for a perpendicular crack.

This paper does not concern the enhancement in toughness as opposed to the
title. Rather, we proceed to study a case of parallel crack with the fixed
grip condition from {\em the continuum view} (as for non-continuum treatment
on laminated composites, see for example, \cite{Leguillon}). As we see
later, this coarse-grained treatment again leads to a reasonably simplified
picture of the problem. To describe the fields near the tip, we shall find
that an extra term is required in our elastic energy. This term introduces a
new length scale $\lambda $, as we see below, and the extra term becomes
important near the tip ($x\ll \lambda $, where $x$ is the distance from the
tip). This term is suggested previously concerning some smectic liquid
crystal \cite{LQ} and a {\em lenticular} tip form is predicted from scaling
arguments. \cite{PGG90} Here, we obtain an analytical solution to the stress
and deformation distribution near the tip; the tip form in this case takes
also the lenticular form. The appearance of a {\em parabolic} differential
equation for the deformation field due to the extra term is another feature
of this paper. We also obtain an analytical solution for $x\gg \lambda $,
which allows us to estimate the fracture energy; there is no energy
enhancement within our treatment.

The Young modulus of the hard and soft layer are denoted $E_{h}$ and $E_{s}$%
, respectively, where 
\[
E_{s}=\varepsilon E_{h}. 
\]%
In the following, we consider the case of parallel fractures under the plane
strain condition, i.e. $e_{zx}=e_{zy}=e_{zz}=0$ (Fig. 1). Here, $e_{ij}$ is
the strain derived from the displacement field $u_{i}$, i.e. 
\[
e_{ij}=\frac{1}{2}\left( \frac{\partial u_{i}}{\partial x_{j}}+\frac{%
\partial u_{j}}{\partial x_{i}}\right) , 
\]%
where $(x_{1},x_{2},x_{3})\equiv (x,y,z).$

The ensuing analysis is based on the conditions appropriate for nacre: 
\begin{eqnarray*}
\varepsilon &\ll &1, \\
d_{s} &\ll &d_{h}, \\
\epsilon &=&\varepsilon d/d_{s}\ll 1,
\end{eqnarray*}%
where, $d=d_{s}+d_{h}\simeq d_{h}$.

\subsection{Elastic energy for the nacre in the thin layer limit}

Assuming that the thickness of layers are thin in the sense that we can
neglect the stress change over a few layers, we can introduce {\it %
macroscopic strain field} to have the following elastic energy \cite{KOPGG} 
\begin{equation}
f=\frac{E}{2(1-\nu ^{2})}e_{xx}^{2}+\frac{E_{0}}{2}e_{yy}^{2}+\frac{E_{0}}{%
1+\nu }e_{xy}^{2}+\frac{\nu E_{0}}{1-\nu }e_{xx}e_{yy},  \label{energy}
\end{equation}%
where 
\begin{eqnarray*}
E &=&E_{h}, \\
E_{0} &=&\epsilon E_{h}.
\end{eqnarray*}%
We have assumed that the Poisson ratio is the same for both layers for
simplicity (even if $E_{s}\gg E_{h}$, or $\varepsilon \ll 1,$ $\nu _{s}$ and 
$\nu _{h}$ are typically of the same order). We have checked (for the
perpendicular crack) the stress distribution resulting from this energy
exhibits small change over a few layers, which justifies the above
assumption.

The strain-stress relation results from this energy by the relation $%
\sigma_{ij}=\partial f/\partial e_{ij}$: 
\begin{mathletters}
\begin{eqnarray}
\sigma _{xx} &=&\frac{E}{1-\nu ^{2}}e_{xx}+\frac{\nu }{1-\nu }E_{0}e_{yy},
\label{stxx} \\
\sigma _{yy} &=&E_{0}\left( e_{yy}+\frac{\nu }{1-\nu }e_{xx}\right) ,
\label{styy} \\
\sigma _{xy} &=&\frac{E_{0}}{1+\nu }e_{xy}.  \label{stxy}
\end{eqnarray}

\subsection{Parallel fracture: inclusion of the bending effect}

Near the fracture tip, the stress gradient is large; we may need to include
the stress changes within a single layer in some cases. In the case of
parallel fracture, the bending, which originates from the non-uniform
stress, is important as we see below. The correction due to the bending
effect is given by (neglecting the bending energy of the soft layer) 
\end{mathletters}
\[
f_{B}=\frac{B_{h}}{2d}\left( \frac{\partial ^{2}u_{y}}{\partial x^{2}}%
\right) ^{2}\equiv \frac{K}{2}\left( \frac{\partial ^{2}u_{y}}{\partial x^{2}%
}\right) ^{2}, 
\]%
where the bending moduli is given by \cite{Landau} 
\[
B_{h}=\frac{E_{h}d_{h}^{3}}{12(1-\nu ^{2})}. 
\]%
The total energy is then given by 
\begin{equation}
f_{T}=f+\frac{K}{2}\left( \frac{\partial ^{2}u_{y}}{\partial x^{2}}\right)
^{2}.  \label{elT}
\end{equation}%
As we see later, in the case of parallel fracture, the first term in $f$ in
Eq. (\ref{energy}) becomes negligible; the remaining terms in $f$ are all
associated with the weak modulus $E_{s}$ while the bending term is
associated with the strong modulus $E_{h}$. This suggests the possibility
that {\em this higher-order derivative term make a significant contribution
to the energy in the parallel fracture.}

In accordance with the introduction of the bending term, we have another
length scale $\lambda $ as announced: 
\[
\lambda ^{2}=\frac{K}{E_{0}}\sim \frac{d^{2}}{\epsilon }. 
\]

\section{Scaling prediction}

As we see below, at large scale where $\lambda \ll x$, we have $u_{x}\ll
u_{y}$, and the elastic energy is simplified to 
\begin{equation}
f=\frac{E_{0}}{2}\left( \frac{\partial u_{y}}{\partial y}\right) ^{2}+\frac{%
E_{0}}{4\left( 1+\nu \right) }\left( \frac{\partial u_{y}}{\partial x}%
\right) ^{2}.  \label{lscale}
\end{equation}

At small scale $\lambda \gg x$, we also have $u_{x}\ll u_{y}$ and the
elastic energy is simplified to 
\begin{equation}
f=\frac{E_{0}}{2}\left( \frac{\partial u_{y}}{\partial y}\right) ^{2}+\frac{K%
}{2}\left( \frac{\partial ^{2}u_{y}}{\partial x^{2}}\right) ^{2}.
\label{sscale}
\end{equation}%
This energy in Eq. (\ref{sscale}) has been discussed for smectic liquid
crystals. \cite{LQ,PGG90} By considering{\em \ energy balance as in the
Griffith theory}, \cite{Griffith} it is predicted that the system exhibits a
''lenticular fracture,'' where the opening angle $\theta _{0}$ is {\em finite%
}, i.e.,%
\begin{equation}
\theta _{0}^{2}\sim \frac{G_{0}}{E_{0}\lambda },  \label{th}
\end{equation}%
where $G_{0}(=2\gamma _{0})$ is a thermodynamic separation energy for the
soft medium (where $\gamma _{0}$ is the surface energy of the organic
substance). Note here this angle is small ($\theta _{0}\ll 1$) because $%
G_{0}/E_{0}\sim a$ , where $a$ is a molecular length; thus we see the angle
is small ($\theta _{0}^{2}\sim a/\lambda )$. (Here, the relation, $%
G_{0}/E_{0}\sim a,$ results from the following argument. The scaling
expression for the elastic modulus for polymer melt is given by $E\sim
cT/N_{e}$ where $c\sim 1/a^{3}$ and $N_{e}$ $\sim 100$, \cite{Polymer} while
the surface tension is typically given by $\gamma =T/a^{2}$. \cite{Cap}
Thus, we have $E\sim \gamma /a$.)

In the inner (small scale) region, the deformation of the free surface is
given by%
\[
u_{s}\equiv u_{y}(x,y=0)=\theta _{0}x. 
\]

In the outer (large scale) region where the bending term is neglected, we
expect the usual singular behavior for the stress and deformation
distribution because the deformation field satisfies the two-dimensional
Laplace equation resulting from the energy in Eq. (\ref{lscale});%
\begin{eqnarray*}
\sigma _{s} &\equiv &\sigma _{yy}(x,y=0)=\frac{K_{I}}{\sqrt{-x}} \\
u_{s} &=&\frac{K_{I}}{E_{0}}\sqrt{x}.
\end{eqnarray*}

Requiring the matching of the deformation field at $x=\lambda $, we have%
\[
K_{I}^{2}=E_{0}^{2}\theta _{0}^{2}\lambda =G_{0}E_{0}. 
\]%
This shows that the stress intensity factor is the same order with that for
a pure organic system; there is no enhancement of the separation energy.

$K_{I}$\ usually depends on the remote tensile stress and, at the moment of
fracture (where $\sigma _{\infty }=\sigma _{Failure}$), the square of this
coincides with the elastic constant multiplied by the fracture energy (e.g. $%
K_{I}^{2}=\mu G$).{\bf \ }In the above, it seems that $K_{I}$\ does not
depend on the remote stress from the beginning. This is because the
expression (\ref{th}) is derived from an energy-balance consideration which
requires the condition ''at the moment of fracture.''

\section{Simplification of the energy at large and small scales}

\bigskip At equilibrium, by requiring $\delta F=F(u_{x}+\delta
u_{x})-F(u_{x})=0$ for any displacement $\delta u_{x}$, where%
\begin{equation}
F=\int d{\bf r}\left( \frac{E}{2(1-\nu ^{2})}e_{xx}^{2}+\frac{E_{0}}{2}%
e_{yy}^{2}+\frac{E_{0}}{1+\nu }e_{xy}^{2}+\frac{\nu E_{0}}{1-\nu }%
e_{xx}e_{yy}+\frac{K}{2}\left( \frac{\partial ^{2}u_{y}}{\partial x^{2}}%
\right) ^{2}\right) ,  \label{F}
\end{equation}%
we have%
\[
\frac{\partial ^{2}u_{x}}{\partial x^{2}}+\epsilon \alpha \frac{\partial
^{2}u_{x}}{\partial y^{2}}+\epsilon \beta \frac{\partial ^{2}u_{y}}{\partial
x\partial y}=0, 
\]%
where%
\[
\alpha =\frac{1}{2\left( 1+\nu \right) },\beta =\frac{1}{2}\left( \frac{1}{%
1+\nu }+\frac{2\nu }{1-\nu }\right) 
\]

For the displacement of $u_{y}$, we have

\[
\frac{\partial ^{2}u_{y}}{\partial y^{2}}+\alpha \frac{\partial ^{2}u_{y}}{%
\partial x^{2}}+\beta \frac{\partial ^{2}u_{x}}{\partial x\partial y}%
-\lambda ^{2}\frac{\partial ^{4}u_{y}}{\partial x^{4}}=0. 
\]

Let us try to find a solution of the form%
\begin{eqnarray*}
u_{x} &=&u\exp (ipx)\exp (iqy) \\
u_{y} &=&v\exp (ipx)\exp (iqy)
\end{eqnarray*}%
and put them into the above two equations. We have%
\[
\left( 
\begin{array}{cc}
p^{2}+\epsilon \alpha q^{2} & \epsilon \beta pq \\ 
\beta pq & q^{2}+(\alpha +\lambda ^{2}p^{2})p^{2}%
\end{array}%
\right) \left( 
\begin{array}{c}
u \\ 
v%
\end{array}%
\right) =\left( 
\begin{array}{c}
0 \\ 
0%
\end{array}%
\right) . 
\]

Requiring $(u,v)\neq (0,0)$, with $\alpha ^{\prime }=\alpha ^{2}-\beta ^{2}$%
, we have 
\[
\epsilon \alpha q^{4}+\left( 1+\epsilon \left( \alpha ^{\prime }+\alpha
\lambda ^{2}p^{2}\right) \right) p^{2}q^{2}+\allowbreak \left( \alpha
+\lambda ^{2}p^{2}\right) p^{4}=0. 
\]%
Since we concern the region, $p\ll 1/d$, for a continuum theory, from the
relation $\lambda ^{2}\sim d^{2}/\epsilon $ we have $\epsilon \lambda
^{2}p^{2}\ll 1$ and $\epsilon \lambda p\ll 1$. Then, we arrive at the two
solutions,%
\[
q^{2}=-p^{2}(\alpha +\lambda ^{2}p^{2})+O(\epsilon ),-\frac{p^{2}}{\epsilon
\alpha }\left( 1+O(\epsilon )\right) 
\]%
The minus sign on the right-hand side indicates that the solution is a
damping function in the $y$ direction and some trigonometric function in the 
$x$-direction.

From the second solution, we have $v\sim \sqrt{\epsilon }u$. Since we seek
the solution with $v>u$ in our boundary condition, we chose the first, i.e., 
$q^{2}=-p^{2}(\alpha +\lambda ^{2}p^{2})$. Thus, at large scale ($\lambda
p\ll 1$), we have $q^{2}=-p^{2}\alpha $, which leads to $u\sim \epsilon v$.
These facts allow us to reduce Eq. (\ref{F}) to the form announced in Eq. (%
\ref{lscale}). At small scale ($\lambda p\gg 1$), we have $q^{2}=-\lambda
^{2}p^{4}$ and, thus, $u\sim \epsilon \lambda pv$, which allows Eq. (\ref{F}%
) to be reduced to the form in Eq. (\ref{sscale}).

Unlike the perpendicular case, \cite{KOPGG} the energy is not simplified\
into a single form over all length scale in the present case. Instead, we
observe the simplification of the energy at both the large and small scales,
i.e., $\lambda p\ll 1$ and $\lambda p\gg 1.$ We note here that $\lambda $ is
rather {\it large} compared with the layer thickness, i.e., $\lambda \sim d/%
\sqrt{\epsilon }$. The ensuing analysis is based on the condition, $d\ll
\lambda \ll L$, where $2L$ is the $y$ dimension of the sample.

\section{Large scale solution}

At equilibrium, by minimizing the energy given in Eq. (\ref{lscale}), we
have 
\[
\left( \frac{\partial ^{2}}{\partial \overline{x}^{2}}+\frac{\partial ^{2}}{%
\partial y^{2}}\right) u_{y}=0, 
\]%
where 
\[
\overline{x}=\sqrt{2(1+\nu )}x. 
\]%
When the crack tip is located at the origin of the $x$-axis the appropriate
boundary conditions are, under the fixed condition, as follows: 
\begin{eqnarray*}
u_{y} &=&\pm u_{0}\text{ at }y=\pm L, \\
u_{y} &=&0\text{ for }y=0,\text{ }x\ll 0(x\ll -\lambda ), \\
\frac{\partial u_{y}}{\partial y} &=&0\text{ for }y=0,\text{ }x\gg 0(x\gg
\lambda ).
\end{eqnarray*}%
Note here that these boundary conditions specify the conditions at the
points {\em away from the origin}.{\bf \ }As in the perpendicular case, in
analogy with the boundary problem for a variable condenser, \cite{KOPGG} we
have the displacement; 
\begin{equation}
u_{y}=\frac{2u_{0}}{\pi }%
\mathop{\rm Im}%
\left[ \log \left( e^{i\pi z/(2L)}+\left( e^{i\pi z/L}-1\right)
^{1/2}\right) \right] ,  \label{lu0}
\end{equation}%
with 
\[
z=y+i\left( \overline{x}-\overline{x}_{0}\right) . 
\]%
In the above, the branch of the function $z^{1/2}$ is chosen such that $%
z^{1/2}=r^{1/2}\exp (i\theta /2)$ with $0<\theta <2\pi $ (e.g., $%
(-1)^{1/2}=i $). Here, the remote boundary conditions allow the shift, $%
x_{0}\left( =\overline{x}_{0}/\sqrt{2(1+\nu )}\right) $, in our solution, as
far as $x_{0}\ll L$.

In the following, we consider only the region $y>0$, since the problem is
symmetric with respect to the $x$-axis. When $\left| z\right| \ll L$, we have%
\begin{equation}
u_{y}(x,y)=\frac{2u_{0}}{\pi }%
\mathop{\rm Im}%
\left[ \left( \frac{i\pi z}{L}\right) ^{1/2}\right] .  \label{lu}
\end{equation}

In the vicinity of the origin (but still $x\gg \lambda $), it reduces to a
parabolic form: 
\begin{equation}
u_{y}(x,y=0)=2u_{0}\theta (x)\sqrt{\frac{\overline{x}-\overline{x}_{0}}{\pi L%
}}.  \label{lscdis}
\end{equation}%
where $\theta (x)$ is the Heaviside step function. Here and hereafter, {\em %
the notation }$y=0${\em \ should be understood as the limit where the
variable }$y${\em \ approaches to an positive infinitesimal quantity. }This
is because the consideration for the region $y>0$ is enough due to the
symmetry.

The nonzero component of the stress field is given from $\sigma
_{yy}=E_{0}e_{yy};$ 
\begin{equation}
\sigma _{yy}=\sigma _{\infty }%
\mathop{\rm Re}%
\left[ \frac{e^{i\pi z/(2L)}}{\left( e^{i\pi z/L}-1\right) ^{1/2}}\right] ,
\label{ls}
\end{equation}%
where the remote tensile stress is given by 
\[
\sigma _{\infty }=E_{0}\frac{u_{0}}{L}. 
\]%
When $\left| z\right| \ll L$, we have%
\[
\sigma _{yy}(x,y)=\sigma _{\infty }%
\mathop{\rm Re}%
\left[ \left( \frac{i\pi z}{L}\right) ^{-1/2}\right] , 
\]%
and stress distribution near the tip is given by 
\begin{equation}
\sigma _{yy}(x,y=0)=\sigma _{\infty }\theta (-x)\sqrt{\frac{L}{\left(
-\left( \overline{x}-\overline{x}_{0}\right) \right) \pi }}.  \label{lstress}
\end{equation}%
Note here that these expressions are valid only when $x>\lambda (\sim d/%
\sqrt{\epsilon })$ where $\lambda \ll L$.

\section{Small scale solution}

From Eq. (\ref{sscale}), we have a ''pseudo {\em bi-diffusion}'' equation, 
\begin{equation}
\frac{\partial ^{2}u_{y}}{\partial y^{2}}=\lambda ^{2}\frac{\partial
^{4}u_{y}}{\partial x^{4}}.  \label{diff0}
\end{equation}%
The solution to our problem has to satisfy the following two boundary
conditions at least%
\begin{eqnarray*}
u_{y} &=&0\text{ for }y=0,\text{ }x<0(\left| x\right| \ll \lambda ), \\
\frac{\partial u_{y}}{\partial y} &=&0\text{ for }y=0,\text{ }x>0(x\ll
\lambda ).
\end{eqnarray*}%
In addition, the solution should be able to match with the large scale
solution.

\subsection{General solution for the boundary problem}

The field $u_{y}$ satisfies the (pseudo) {\em bi-diffusion} equation,

\[
\left( \partial _{y}+\lambda \partial _{x}^{2}\right) \left( \partial
_{y}-\lambda \partial _{x}^{2}\right) u_{y}=0. 
\]%
Putting $S_{u}=\left( \partial _{y}-\lambda \partial _{x}^{2}\right) u_{y}$,
we see that $S_{u}$ is a solution to the anti-diffusion equation%
\[
\left( \partial _{y}+\lambda \partial _{x}^{2}\right) S_{u}=0 
\]%
and that $u_{y}$ satisfies the inhomogeneous diffusion equation%
\[
\left( \partial _{y}-\lambda \partial _{x}^{2}\right) u_{y}=S_{u}. 
\]%
The source function $S_{u}$ is a solution of the anti-diffusion equation (a
typical solution is $\exp \left( \frac{x^{2}}{4\lambda y}\right) /\sqrt{4\pi
\lambda y}$) and blows up when $y$ approaches to zero, except that $S_{u}$
is independent of $y$, i.e., $S_{u}=A+Bx$ (this fact can be confirmed later)
where $A$ and $B$ are constants independent of $x$ and $y$.

Since $u_{y}$ satisfies the bi-diffusion equation, $\sigma _{yy}$ also
satisfies the bi-diffusion equation. Putting $S_{\sigma }=\left( \partial
_{y}-\lambda \partial _{x}^{2}\right) \sigma _{yy}$, we have%
\[
\left( \partial _{y}-\lambda \partial _{x}^{2}\right) \sigma _{yy}=S_{\sigma
}. 
\]%
To have a physical solution, we have $S_{u}=A+Bx$ and we conclude that $%
S_{\sigma }=0$ because $S_{\sigma }=E_{0}\partial _{y}S_{u}$. Namely,{\em \ }%
$\sigma _{yy}${\em \ is the solution of the diffusion equation. }As we see
later, the function $S_{u}$ obtained starting from the diffusion equation
for $\sigma _{yy}$ is actually in the form, $S_{u}=A+Bx$. In this way, we
shall obtain the general solution for $u_{y}$.

The general solution for the initial boundary condition (here, ''initial''
corresponds to $y=0$)%
\begin{equation}
\sigma _{yy}(x,y=0)=F(x)  \label{s0}
\end{equation}%
is given as%
\begin{equation}
\sigma _{yy}(x,y)=\frac{1}{\sqrt{4\pi \lambda y}}\int_{-\infty }^{\infty
}F(x^{\prime })\exp \left( -\frac{\left( x-x^{\prime }\right) ^{2}}{4\lambda
y}\right) dx^{\prime }.  \label{1}
\end{equation}%
In the following, we determine {\it the source function} $F(x)$ considering
appropriate boundary conditions. Requiring the boundary condition 
\begin{equation}
\sigma _{yy}(x,y)=0\text{ for }y=0\text{ and }x>0,  \label{b1}
\end{equation}%
we may put%
\begin{equation}
F(x)=\theta (-x)f(x)+\sigma \lambda \delta (x)  \label{sf}
\end{equation}%
where $\delta (x)$ is the Dirac's delta function and $f(x)$ and $\sigma $
are the quantities to be determined. As we see later, the second singular
term in the source function is required to have a smooth matching to the
outer solution. In this way, we have%
\begin{equation}
\sigma _{yy}(x,y)=\frac{1}{\sqrt{4\pi \lambda y}}\int_{-\infty
}^{0}f(x^{\prime })\exp \left( -\frac{\left( x-x^{\prime }\right) ^{2}}{%
4\lambda y}\right) dx^{\prime }+\sigma \sqrt{\frac{\lambda }{4\pi y}}\exp
\left( -\frac{x^{2}}{4\lambda y}\right) .  \label{eq1}
\end{equation}

By integrating Eq. (\ref{eq1}) over $y$, we have ($%
\mathop{\rm erfc}%
\left( x\right) =1-%
\mathop{\rm erf}%
(x)$, $%
\mathop{\rm erf}%
(x)=\frac{2}{\sqrt{\pi }}\int_{0}^{x}e^{-t^{2}}dt$)%
\begin{eqnarray}
E_{0}u_{y}(x,y) &=&\int_{-\infty }^{0}f(x^{\prime })\left( \sqrt{\frac{y}{%
\pi \lambda }}e^{-\frac{\left( x-x^{\prime }\right) ^{2}}{4\lambda y}}-\frac{%
x-x^{\prime }}{2\lambda }%
\mathop{\rm erfc}%
\left( \frac{x-x^{\prime }}{\sqrt{4\lambda y}}\right) \right) dx^{\prime } 
\nonumber \\
&&+\sigma \left( \sqrt{\frac{\lambda y}{\pi }}\exp \left( -\frac{x^{2}}{%
4\lambda y}\right) -\frac{x}{2}%
\mathop{\rm erfc}%
\left( \frac{x}{\sqrt{4\lambda y}}\right) \right) +C(x),  \label{eq2}
\end{eqnarray}%
where we have used the formula given in \cite{formula1,formula2}, in which
formulae the integration constants are chosen so that the result vanishes at 
$x=-\infty $. \noindent In the above, $C(x)$ is the solution of the
bi-diffusion equation independent of $y$, i.e., 
\[
C(x)=c_{0}+c_{1}x+c_{2}x^{2}+c_{3}x^{3}. 
\]

The boundary condition required for $u_{y}(x,y)$ is%
\begin{equation}
u_{y}(x,y)=0\text{ for }y=0\text{ and }x<0.  \label{b2}
\end{equation}%
To determine the function $f(x)$ to accommodate this boundary condition, we
consider $u_{y}(x,y)$ at $y=0$. Noting the relation,%
\begin{eqnarray*}
&&\int_{-\infty }^{0}f(x^{\prime })\frac{x-x^{\prime }}{2\lambda }%
\mathop{\rm erfc}%
\left( \frac{x-x^{\prime }}{2\sqrt{\lambda y}}\right) dx^{\prime } \\
&=&\left( \int_{-\infty }^{x}dx^{\prime }+\int_{x}^{0}dx^{\prime }\right)
f(x^{\prime })\frac{x-x^{\prime }}{2\lambda }%
\mathop{\rm erfc}%
\left( \frac{x-x^{\prime }}{2\sqrt{\lambda y}}\right) ,
\end{eqnarray*}%
where the first term in the right-hand side is zero at $y=0$ and the second
term is nonzero only when $x<0$, we have 
\[
E_{0}u_{y}(x,y=0)=\theta \left( -x\right) \left( \frac{1}{\lambda }%
\int_{0}^{x}f(x^{\prime })\left( x-x^{\prime }\right) dx^{\prime }-\sigma
x\right) +C(x). 
\]%
Here, note the relation $%
\mathop{\rm erfc}%
\left( \left( x-x^{\prime }\right) /\sqrt{4\lambda y}\right) \longrightarrow
2\theta \left( -x\right) $ when $y\rightarrow 0^{+}.$

Since $\theta (-x)=1-\theta (x)$, the boundary condition (\ref{b2}) is
satisfied only when%
\begin{equation}
C(x)=\sigma x-\frac{1}{\lambda }\int_{0}^{x}f(x^{\prime })\left( x-x^{\prime
}\right) dx^{\prime },  \label{cx}
\end{equation}%
to have%
\[
E_{0}u_{y}(x,y=0)=\theta \left( x\right) \left( \sigma x-\frac{1}{\lambda }%
\int_{0}^{x}f(x^{\prime })\left( x-x^{\prime }\right) dx^{\prime }\right) . 
\]

The condition (\ref{cx}) is strong in the sense that $C(x)$ is the
polynomial of $x$ with the order lower than the fourth and is independent of 
$y$; $f(x)$ is at most linear in $x$,%
\begin{equation}
f(x)=a_{0}+a_{1}x.  \label{fx}
\end{equation}%
In this way, we obtain%
\begin{equation}
E_{0}u_{y}(x,y=0)=\theta \left( x\right) \left( \sigma x-\frac{1}{\lambda }%
\left( \frac{1}{2}a_{0}x^{2}+\allowbreak \frac{1}{6}a_{1}x^{3}\right) \right)
\label{uy0}
\end{equation}%
and from Eqs. (\ref{s0}) and (\ref{sf}), we have (for $x\neq 0$)%
\begin{equation}
\sigma _{yy}(x,y=0)=\theta (-x)(a_{0}+a_{1}x)  \label{sy0}
\end{equation}

\section{Matching between inner and outer solutions}

Eq. (\ref{uy0}) contains the three unknown constants, $a_{0},a_{1},$ and $%
\sigma ,$ while the outer solution the one unknown, $x_{0}$; there are four
unknowns. To determine these constants we require that $u_{y}$ and $\partial
_{x}u_{y}$ match the outer solution at $x=\lambda $ while $\sigma _{yy}$ and 
$\partial _{x}\sigma _{yy}$ match the large-scale solution at $x=-\lambda $.

\subsection{Outer solution}

Introducing $\alpha $ (which is different from $\alpha $ used in Sec. III)%
\[
\alpha ^{2}=\sqrt{2\left( 1+\nu \right) } 
\]%
we have

\begin{eqnarray*}
u_{y}(x,y &=&0)=2u_{0}\alpha \theta (x)\sqrt{\frac{x-x_{0}}{\pi L}} \\
\frac{\partial }{\partial x}u_{y}(x,y &=&0)=\theta (x)\frac{u_{0}\alpha }{%
\sqrt{\pi L\left( x-x_{0}\right) }} \\
\frac{1}{E_{0}}\sigma _{yy}(x,y &=&0)=\theta (-x)\frac{u_{0}}{\alpha \sqrt{%
\pi L\left( -\left( x-x_{0}\right) \right) }} \\
\frac{1}{E_{0}}\frac{\partial }{\partial x}\sigma _{yy}(x,y &=&0)=\theta (-x)%
\frac{u_{0}}{2\alpha \sqrt{\pi L}}\left( -\left( x-x_{0}\right) \right)
^{-3/2}
\end{eqnarray*}

\subsection{Inner solution}

Introducing $a$, $b$ and $u$ (which is different from $u$ in Sec. III)%
\[
a\equiv \frac{\lambda a_{0}}{E_{0}u},b\equiv \frac{a_{1}\lambda ^{2}}{E_{0}u}%
,u\equiv \frac{\sigma \lambda }{E_{0}} 
\]%
we have

\begin{eqnarray*}
u_{y}(x,y &=&0)=u\theta (x)\left( \frac{x}{\lambda }-\left( \frac{a}{%
2\lambda ^{2}}x^{2}+\frac{b}{6\lambda ^{3}}x^{3}\right) \right) \\
\frac{\partial }{\partial x}u_{y}(x,y &=&0)=u\theta (x)\left( \frac{1}{%
\lambda }-\left( \frac{a}{\lambda ^{2}}x+\frac{b}{2\lambda ^{3}}x^{2}\right)
\right) \\
\frac{1}{E_{0}}\sigma _{yy}(x,y &=&0)=\theta (-x)\frac{u}{\lambda }\left( a+%
\frac{b}{\lambda }x\right) \\
\frac{1}{E_{0}}\frac{\partial }{\partial x}\sigma _{yy}(x,y &=&0)=\theta
(-x)u\frac{b}{\lambda ^{2}}
\end{eqnarray*}

\subsection{Matching}

We have four conditions, while we have four unknowns, $a,b,x_{0},$ and $u$,
for the four conditions.

At $x=\lambda :$%
\begin{eqnarray*}
-u\left( \frac{a}{2}+\frac{b}{6}\right) +u &=&2\alpha u_{0}l\sqrt{\gamma } \\
-u\left( \frac{a}{\lambda }+\frac{b}{2\lambda }\right) +u\frac{1}{\lambda }
&=&\frac{\alpha u_{0}l}{\lambda \sqrt{\gamma }}
\end{eqnarray*}%
Here, we have introduced%
\[
l=\sqrt{\frac{\lambda }{\pi L}},\gamma =1-\frac{x_{0}}{\lambda } 
\]

At $x=-\lambda :$%
\begin{eqnarray*}
\frac{u}{\lambda }\left( a-b\right) &=&\frac{u_{0}l}{\alpha \lambda \sqrt{%
\gamma ^{\prime }}} \\
u\frac{b}{\lambda ^{2}} &=&\frac{u_{0}l}{2\alpha \lambda \lambda ^{2}\left( 
\sqrt{\gamma ^{\prime }}\right) ^{3}}
\end{eqnarray*}%
Here, we have introduced%
\[
\gamma ^{\prime }\equiv 1+\frac{x_{0}}{\lambda }, 
\]%
from which we have $-x+x_{0}=\lambda (1+x_{0}/\lambda )=\lambda \gamma
^{\prime }$. With the definition, 
\[
\eta =\frac{u_{0}}{u}l, 
\]%
we have%
\begin{eqnarray}
a &=&4-2\alpha \eta \frac{6\gamma -1}{\sqrt{\gamma }}  \nonumber \\
b &=&-6+6\alpha \eta \frac{4\gamma -1}{\sqrt{\gamma }}  \label{ab}
\end{eqnarray}%
and%
\begin{eqnarray*}
a &=&\eta \frac{2\gamma ^{\prime }+1}{2\alpha (\gamma ^{\prime })^{\frac{3}{2%
}}} \\
b &=&\frac{\eta }{2\alpha (\gamma ^{\prime })^{\frac{3}{2}}}
\end{eqnarray*}%
Thus, we have a set of equations for $\gamma $ and $\eta :$%
\begin{eqnarray}
4-2\alpha \eta \frac{6\gamma -1}{\sqrt{\gamma }} &=&\eta \frac{2\gamma
^{\prime }+1}{2\alpha (\gamma ^{\prime })^{\frac{3}{2}}}  \nonumber \\
-6+6\alpha \eta \frac{4\gamma -1}{\sqrt{\gamma }} &=&\frac{\eta }{2\alpha
(\gamma ^{\prime })^{\frac{3}{2}}}  \label{ge}
\end{eqnarray}%
$\allowbreak $

Numerical solution to this set of equation with $\left| x_{0}\right|
<\lambda $ (or $\left| 1-\gamma \right| <1$) is $\gamma =0.71842,\eta
=0.38752$ for $\nu =0$ and $\eta =0.378\,70,\gamma =0.6656$ for $\nu =1/2$.

Approximate but analytical expressions for them can be obtained as follows.
Introducing $\eta ^{\prime }\equiv 1/\eta $ and $\delta $ by

\[
\delta =x_{0}/\lambda ,\gamma =1-\delta ,\gamma ^{\prime }=1+\delta 
\]%
and, assuming, $\delta \ll 1$, the set of equation can be reduced to%
\begin{eqnarray*}
4\eta ^{\prime }-2\alpha \allowbreak \left( 5-\frac{7}{2}\delta \right) &=&%
\frac{6-5\delta }{4\alpha } \\
-6\eta ^{\prime }+6\alpha \left( 3-\frac{5}{2}\delta \right) &=&\frac{%
2-3\delta }{4\alpha }
\end{eqnarray*}%
Solving this equation we have an approximate solution:%
\begin{eqnarray*}
\delta &=&\frac{2\left( 12\alpha ^{2}-11\right) }{3\left( 12\alpha
^{2}-7\right) } \\
\eta ^{\prime } &=&\frac{48\alpha ^{4}-8\alpha ^{2}-1}{3\left( 12\alpha
^{2}-7\right) \alpha }
\end{eqnarray*}%
For $\alpha =\sqrt{2(1+\nu )}$ with $0<\nu <1/2$, $\delta \simeq 0.4$ which
barely justifies the assumption, $\delta \ll 1$, while $\eta ^{\prime
}\simeq 2.4$. In this way, we confirm that $\gamma ${\em \ and }$\eta ${\em %
\ are both positive quantity of the order of unity }for any value of $\nu $ (%
$0<\nu <1/2$).

\section{Overall solution}

For $\left| x\right| <\lambda $, from Eqs. (\ref{eq2}), (\ref{cx}), and (\ref%
{fx}) we have 
\begin{eqnarray}
\frac{u_{y}(x,y)}{u} &=&-\frac{1}{2}\left( \frac{a}{2}X^{2}+\frac{b}{6}%
X^{3}\right) \left( 1+%
\mathop{\rm erf}%
\left( \frac{X}{\sqrt{4Y}}\right) \right)  \label{ovall} \\
&&+\frac{1}{2}Y\left( a+bX\right) 
\mathop{\rm erfc}%
\left( \frac{X}{\sqrt{4Y}}\right) -\left( \frac{2}{3}bY+\frac{\frac{a}{2}%
X^{2}+\frac{b}{6}X^{3}}{X}\right) \sqrt{\frac{Y}{\pi }}e^{-\frac{X^{2}}{4Y}}
\nonumber
\end{eqnarray}%
and from Eqs. (\ref{eq1}) and (\ref{fx}) we have

\[
\frac{\sigma _{yy}(x,y)\lambda }{E_{0}}=\frac{1}{2}\left( a+bX\right) 
\mathop{\rm erfc}%
\left( \frac{X}{\sqrt{4Y}}\right) -b\sqrt{\frac{Y}{\pi }}e^{-\frac{X^{2}}{4Y}%
}\allowbreak 
\]%
where $a$ $(\sim 0.4)$ and $b$ $(\sim 0.1)$ are given by Eq. (\ref{ab}) and $%
u=u_{0}l/\eta $ $(\sim 2.6u_{0}l)$ with $\gamma $ and $\eta $ given by Eq. (%
\ref{ge}). Here, $X=x/\lambda $ and $Y=y/\lambda $.

For $\left| x\right| >\lambda $, $u_{y}$ and $\sigma _{yy}$ are given by
Eqs. (\ref{lu0}) and (\ref{ls}) where $x_{0}=(1-\gamma )\lambda $ $(\sim
0.3\lambda )$.

As expected and clear from (\ref{ovall}), these expression are only valid
for $y\ll \lambda $ (at small scale, we have $q^{2}=-\lambda ^{2}p^{4}$ with 
$\lambda p\gg 1$, from which we have $q^{2}\gg p^{2}\gg 1/\lambda $).

If we operate $\left( \partial _{y}-\lambda \partial _{x}^{2}\right) $ to
the right-hand side of Eq. (\ref{ovall}) we obtain $a+bX$, which corresponds
to $S_{u}$ in the previous section and thus confirm the statement mentioned
there.

A typical shape of the crack tip and stress distribution around the surface $%
(0<y\ll \lambda )$ are given in the plots in Fig. 2 with the parameters
given in Discussion. The solution becomes precise only when $x\ll \lambda $
or $x\gg \lambda $. Near the tip ($x\ll \lambda $), the tip is {\em %
lenticular} ($\sim x$) while, away from the tip ($x\gg \lambda $), the
deformation field recovers the ordinary square-root law ($\sim \sqrt{x}$) . %
\cite{Rice,Anderson,Lawn}

\section{Discussion}

Since we have the usual singular behavior for the stress and stress fields
we have%
\begin{equation}
G_{0}\sim \left( \sigma u\right) _{\text{large scale}}\sim \sigma _{\infty
}u_{0}.  \label{G}
\end{equation}%
On the other hand, from the Griffith's energy balance, we have $\gamma
_{0}\sim L\sigma _{\infty }^{2}/E_{0}$. \cite{PGGKO} Thus, the fracture
energy is the order of $\gamma _{0}$ as predicted by the scaling
consideration.

Since we have%
\[
\overline{\theta }_{0}(x)\equiv \frac{\partial }{\partial x}%
u_{y}(x,y=0)=u\theta (x)\left( \frac{1}{\lambda }-\left( \frac{a}{\lambda
^{2}}x+\frac{b}{2\lambda ^{3}}x^{2}\right) \right) 
\]%
the crack tip angle at $x=0$ is%
\[
\theta _{0}\equiv \overline{\theta }_{0}(0)=\frac{u_{0}l}{\eta \lambda }\sim 
\frac{u_{0}}{\sqrt{L\lambda }} 
\]%
By rewriting the last expression with the aide of (\ref{G}), we have%
\[
\theta _{0}\sim \sqrt{\frac{\gamma _{0}}{E_{0}\lambda }} 
\]%
as predicted in Eq. (\ref{th}).

Typical parameters for nacre are as follows:%
\begin{eqnarray*}
d &=&1{\rm \mu m} \\
L &=&1%
\mathop{\rm cm}
\\
\epsilon &=&1/250
\end{eqnarray*}%
For this parameter we have%
\begin{eqnarray*}
\lambda &=&250{\rm \mu m} \\
L/\lambda &=&40 \\
d/\lambda &=&1/250
\end{eqnarray*}%
These numerical values allow us the continuum description with length scales 
$\lambda $ and $L$ since the relation, $d\ll \lambda \ll L$, holds for these
values.

If we did not include the last term in Eq. (\ref{sf}), we would not have a
linear term in the deformation field and it would contradict with the
scaling expectation (in addition, we could not make a smooth matching as we
see below); we would have

\begin{eqnarray}
u_{y} &=&-u\left( \frac{a}{2\lambda ^{2}}x^{2}+\frac{b}{6\lambda ^{3}}%
x^{3}\right) \theta (x)  \nonumber \\
\frac{1}{E_{0}}\sigma _{yy} &=&\frac{u}{\lambda }\left( a+\frac{b}{\lambda }%
x\right) \theta (-x)  \label{eq}
\end{eqnarray}%
Here, we have only two unknowns; $a^{\prime }\equiv ua$ and $b^{\prime
}\equiv ub$. Let us assume first that $a^{\prime }\neq 0$. Then, the first
equation implies $a^{\prime }<0$ for a positive $u_{y}$ around the origin ($%
x=0$), while the second equation implies $a^{\prime }>0$ for a positive $%
\sigma _{yy}$ (We have assumed $u>0$). Thus, for consistency, we have $%
a^{\prime }=0$.

Thus, we have%
\begin{eqnarray*}
u_{y}(x,y) &=&-\frac{b^{\prime }}{6\lambda ^{3}}x^{3}\theta (x) \\
\frac{1}{E_{0}}\sigma _{yy} &=&\frac{b^{\prime }}{\lambda ^{2}}x\theta (-x)
\end{eqnarray*}%
For positive $u_{y}$ and $\sigma _{yy}$, we have $b^{\prime }<0.$ By
introducing the shift $x_{0}$ for the outer solution as before, the matching
of the both fields themselves is possible. However, since the inner solution 
$u_{y}$ $(\sim x^{3})$ is concave (positive curvature) while the outer
solution $(\sim \sqrt{x})$ convex (negative curvature), the smooth matching
is impossible. For a similar reason, the smooth matching of the stress is
also impossible and the stress field at $x=0$ tends to zero. These facts
suggest the need for the last term in Eq. (\ref{sf}).

\section{Conclusion}

In this paper, we have developed a continuum theory for a nanoscale layered
structure with a parallel crack. We indicated the importance of the bending
term near the crack, which leads to a parabolic differential equation and,
thereby, a lenticular tip form. From the coarse-grained view, we have a
simplified picture with analytical expressions; the deformation field is
linear function of $x$ (the distance from the tip) near the tip and goes
back to the usual square-root function of $x$ away from the tip. As a
result, the fracture energy of nacre against the parallel crack is of the
same order as the soft organic material; there is no enhancement in
toughness within our simplified view.

\acknowledgements

The author is grateful to Pierre-Gilles de Gennes for a number of essential
discussions at the early stage of this work. He would like to express his
sincere gratitude to all the members of de Gennes' group at Coll\`{e}ge de
France for warm hospitality during his stay in Paris. He also thanks to Elie
Rapha\"{e}l for discussions. This work is supported by Joint Research
Project between Japan Society for the Promotion of Science (JSPS) and Centre
National de la Recherche Scientifique (CNRS).

\begin{figure}[tbp]
\caption{Nacre structure: the (inorganic) hard layer thickness $d_{h}$ is of
order of micrometer while the (organic) soft layer thickness $d_{s}$
nanometer. The $y$-axis is perpendicular to layers and the sample is long in
the $z$-direction. The cracks in the $y-z$ plane and in the $x-z$ plane are
called the perpendicular and the parallel fractures, respectively.}
\label{f1}
\end{figure}

\begin{figure}[tbp]
\caption{Crack shape (a) and stress distribution (b) for a parallel crack.
The bold line corresponds to the overall solution while the fine line and
the broken line correspond to the large-scale solution and the small-scale
solution, respectively. The overall solution becomes exact only when $x\ll 
\protect\lambda $ or $x\gg \protect\lambda $. Near the tip ($x\ll \protect%
\lambda $), the tip is a linear function ($\sim x$) while, away from the tip
($x\gg \protect\lambda $), the deformation field recovers the ordinary
square-root law ($\sim x^{1/2}$) . The parameters used for the numerical
calculation are $d=1{\rm \protect\mu m}$, $L=1%
\mathop{\rm cm}%
$, and $\protect\lambda =250{\rm \protect\mu m}$ where $\protect\epsilon %
=1/250$, $L/\protect\lambda =40$, and $d/\protect\lambda =1/250$ (See
Discussion)}
\label{f2}
\end{figure}


\begin{references}
\bibitem{Kelly} T. Kelly and B. Clyne, Physics Today {\bf 52}, 37 (1999).

\bibitem{Clegg} William J. Clegg, Science {\bf 286}, 1097 (1999).

\bibitem{Composites} Ceramic matrix composites: components, preparation,
microstructure and properties, ed. by R. Naslain and B. Harris (Elsevier
Applied Science, 1989); High-temperature ceramic-matrix composites (I, II),
ed. by A. G. Evans and R. Naslain (American Ceramic Society, 1994,95);S. T.
Mileiko, Metal and Ceramic Based Composites (Elsevier Science Ltd., 1997);
Krishan Kumar Chawla and Krishnan Kumar Chawla, Composite Materials: Science
and Engineering (Springer Verlag, 1998); A. Kelly, N. H. Macmillan, Strong
Solids (Oxford U. P., Oxford, 1986); D. Hull and T. W. Clyne, An
Introduction to Composite Materials (Cambridge U. P., Cambridge, 1993).

\bibitem{KOPGG} K. Okumura and P.-G. de Gennes, Eur. Phys. J. E 4, 121
(2001).

\bibitem{PGGKO} P.-G. de Gennes and K. Okumura, C. R. Acad. Sci. Paris t.1,
Ser. IV, 257 (2000).

\bibitem{Currey} J. D. Currey, Proc. R. Soc. Lond. B{\bf 196}, 443 (1977).

\bibitem{Jackson} A. P. Jackson, J. F. V. Vincent, and R. M. Turner, Proc.
R. Soc. Lond. B{\bf 234}, 415 (1988).

\bibitem{Sarikaya} M. Sarikaya, J. Liu, and I. A. Aksay, {\it Biomimetics:
Design and Processing of Materials}, pp.35-90 (AIP Press, New York, 1995).

\bibitem{Rao} M. P. Rao, A. J. S\'{a}nchez-Herencia, G. E. Beltz, R. M.
McMeeking, and F. F. Lange, Science {\bf 286}, 102 (1999).

\bibitem{Leguillon} D. Leguillon and E. Sanchez-Palencia, Fracture in
Heterogeneous Materials, in New Advances in Computational Structural
Mechanics, Edited by P. Ladev\`{e}ze and O.C. Zienkiewicz, p.423 (Elsevier,
1992); Int. J. Fracture 99, 25 (1999).

\bibitem{LQ} P. G. de Gennes, The Physics of Liquid Crystal, Clarendon Press
(Oxford, London, 1974).

\bibitem{PGG90} P. G. de Gennes, Europhys. Lett., 13 (8), 709 (1990).

\bibitem{Landau} L. D. Landau and E. M. Lifshitz, Theory of Elasticity 3rd
Ed. Butterworth-Heinemann (1995).

\bibitem{Griffith} Griffith A.A., Phil. Trans. Roy. Soc. London A 221 (1920)
163--180.

\bibitem{Polymer} P. G. de Gennes, Scaling concepts in Polymer Physics,
Cornell University Press (Ithaca, 1979).

\bibitem{Cap} F. Brochard-Wyart, D. Qu\'{e}r\'{e}, and P.G. de Gennes, Le
monde des gouttes, bulles, perles et ondes (2001, to be published).

\bibitem{formula1} $\int \left( \frac{1}{\sqrt{4\pi \lambda y}}\exp \left( -%
\frac{\left( x-x^{\prime }\right) ^{2}}{4\lambda y}\right) \right) dy=\sqrt{%
\frac{y}{\pi \lambda }}e^{-\frac{\left( x-x^{\prime }\right) ^{2}}{4\lambda y%
}}-\frac{x-x^{\prime }}{2\lambda }%
\mathop{\rm erfc}%
\left( \frac{x-x^{\prime }}{\sqrt{4\lambda y}}\right) $

\bibitem{formula2} $\int \left( \sqrt{\frac{\lambda }{4\pi y}}\exp \left( -%
\frac{x^{2}}{4\lambda y}\right) \right) dy=\sqrt{\frac{\lambda y}{\pi }}\exp
\left( -\frac{x^{2}}{4\lambda y}\right) -\frac{x}{2}%
\mathop{\rm erfc}%
\left( \frac{x}{\sqrt{4\lambda y}}\right) $

\bibitem{Rice} J. R. Rice, {\it Fracture: an Advanced Treatise}, Ed. by H.
Liebowitz, vol. II, pp. 191-311 (Academic Press, New York, 1968).

\bibitem{Anderson} T. L. Anderson, {\it Fracture Mechanics-Fundamentals and
Applications} (CRC Press, Boca Raton, 1991).

\bibitem{Lawn} B. Lawn, Fracture of Brittle Solids, 2nd Ed. (Cambridge Univ.
Press, New York, 1993).
\end{references}
\end{document}